\documentstyle[12pt]{article}

\def\hybrid{\topmargin -20pt    \oddsidemargin 0pt
        \headheight 0pt \headsep 0pt
        \textwidth 6.25in       % A4 paper
        \textheight 9.5in       % A4 paper
        \marginparwidth .875in
        \parskip 5pt plus 1pt   \jot = 1.5ex}
\def\cQ{{\cal Q}}
\def\cG{{\cal G}}
\def\cL{{\cal L}}
\def\cH{{\cal H}}
\def\ket#1{|{#1}\rangle}
\def\noi{\noindent}
\def\half{{1\over2}}
\def\baselinestretch{1.2}

\catcode`\@=11

\def\marginnote#1{}
\def\draftlabel#1{{\@bsphack\if@filesw {\let\thepage\relax
   \xdef\@gtempa{\write\@auxout{\string
      \newlabel{#1}{{\@currentlabel}{\thepage}}}}}\@gtempa
   \if@nobreak \ifvmode\nobreak\fi\fi\fi\@esphack}
        \gdef\@eqnlabel{#1}}
\def\@eqnlabel{}
\def\@vacuum{}
\def\draftmarginnote#1{\marginpar{\raggedright\scriptsize\tt#1}}

\def\draft{\oddsidemargin -.2truein
        \def\@oddfoot{\sl preliminary draft \hfil
        \rm\thepage\hfil\sl\today\quad\militarytime}
        \let\@evenfoot\@oddfoot \overfullrule 3pt
        \let\label=\draftlabel
        \let\marginnote=\draftmarginnote
   \def\@eqnnum{(\theequation)\rlap{\kern\marginparsep\tt\@eqnlabel}%
\global\let\@eqnlabel\@vacuum}  }

\def\preprint{\twocolumn\sloppy\flushbottom\parindent 2em
        \leftmargini 2em\leftmarginv .5em\leftmarginvi .5em
        \oddsidemargin -.5in    \evensidemargin -.5in
        \columnsep .4in \footheight 0pt
        \textwidth 10.in        \topmargin  -.4in
        \headheight 12pt \topskip .4in
88      \textheight 6.9in \footskip 0pt
        \def\@oddhead{\thepage\hfil\addtocounter{page}{1}\thepage}
        \let\@evenhead\@oddhead \def\@oddfoot{} \def\@evenfoot{} }

\def\numberbysection{\@addtoreset{equation}{section}
        \def\theequation{\thesection.\arabic{equation}}}

\def\underline#1{\relax\ifmmode\@@underline#1\else
        $\@@underline{\hbox{#1}}$\relax\fi}

\def\titlepage{\@restonecolfalse\if@twocolumn\@restonecoltrue
\onecolumn
     \else \newpage \fi \thispagestyle{empty}\c@page\z@
        \def\thefootnote{\fnsymbol{footnote}} }

\def\endtitlepage{\if@restonecol\twocolumn \else \newpage \fi
        \def\thefootnote{\arabic{footnote}}
        \setcounter{footnote}{0}}  %\c@footnote\z@ }

\catcode`@=12
\relax

\def\figcap{\section*{Figure Captions\markboth
        {FIGURECAPTIONS}{FIGURECAPTIONS}}\list
        {Figure \arabic{enumi}:\hfill}{\settowidth\labelwidth{Figure
999:}
        \leftmargin\labelwidth
        \advance\leftmargin\labelsep\usecounter{enumi}}}
 \relax
\def\tablecap{\section*{Table Captions\markboth
        {TABLECAPTIONS}{TABLECAPTIONS}}\list
        {Table \arabic{enumi}:\hfill}{\settowidth\labelwidth{Table
999:}
        \leftmargin\labelwidth
        \advance\leftmargin\labelsep\usecounter{enumi}}}
 \relax
\def\reflist{\section*{References\markboth
        {REFLIST}{REFLIST}}\list
        {[\arabic{enumi}]\hfill}{\settowidth\labelwidth{[999]}
        \leftmargin\labelwidth
        \advance\leftmargin\labelsep\usecounter{enumi}}}
 \relax
\makeatletter
\newcounter{pubctr}
\def\publist{\@ifnextchar[{\@publist}{\@@publist}}
\def\@publist[#1]{\list
        {[\arabic{pubctr}]\hfill}{\settowidth\labelwidth{[999]}
        \leftmargin\labelwidth
        \advance\leftmargin\labelsep
        \@nmbrlisttrue\def\@listctr{pubctr}
        \setcounter{pubctr}{#1}\addtocounter{pubctr}{-1}}}
\def\@@publist{\list
        {[\arabic{pubctr}]\hfill}{\settowidth\labelwidth{[999]}
        \leftmargin\labelwidth
        \advance\leftmargin\labelsep
        \@nmbrlisttrue\def\@listctr{pubctr}}}
 \relax
\makeatother
\newskip\humongous \humongous=0pt plus 1000pt minus 1000pt

\newif\ifdtup

\font\Scbig=cmss10 scaled\magstep1
\font\Scscr=cmss8 scaled\magstep1
\font\Scscrscr=cmss8
\newfam\Scfam
\textfont\Scfam=\Scbig
\scriptfont\Scfam=\Scscr
\scriptscriptfont\Scfam=\Scscrscr
\def\Sc{\fam\Scfam}

\relax
\hybrid
\def\lvm{\leavevmode\hbox to\parindent{\hfill}}

\def\thefootnote{\fnsymbol{footnote}}
\def\BE{\begin{equation}}
\def\EE{\end{equation}}
\def\BA{\begin{eqnarray}}
\def\EA{\end{eqnarray}}

\def\th{\theta}

\def\D{\Delta}

\def\tt{\bar\tau}
\def\Gz{\cG_0}
\def\Gn{$\Gz$}
\def\Qz{\cQ_0}
\def\Qn{$\Qz$}

\def\lvm{\leavevmode\hbox to\parindent{\hfill}}
\def\bar{\overline}
\def\req#1{(\ref{#1})}

\def\L{\left}
\def\R{\right}

\def\BE{\begin{equation}}
\def\EE{\end{equation} \vskip 0.30\baselineskip}
\def\BA{\begin{array}}
\def\EA{\end{array}}

\def\noi{\noindent}

\def\frac#1#2{{\textstyle{{#1}\over{#2}}}}
\def\half{{1\over2}}

\def\Kr#1{\delta_{{#1},0}}
\def\ket#1{|{#1}\rangle}

\def\cA{{\cal A}}
\def\cG{{\cal G}}
\def\cH{{\cal H}}

\def\cL{{\cal L}}

\def\cQ{{\cal Q}}

\def\cU{{\cal U}}

\def\open#1{\mbox{{\bf{#1}}}}

\def\oZ{{\open Z}}
\def\o1{{\open 1}}

\def\ctop{{\Sc c}}

\def\kc{\ket\chi}
\def\kca{\ket{\chi^{(1)}}}

\def\ie{{i.e.}}

\newif\ifold \oldtrue \def\new{\oldfalse}
\let\ssection=\section
\def\section{\setcounter{equation}{0}\ssection}

\begin{document}
\renewcommand{\theequation}{\thesection.\arabic{equation}}
\newcommand{\beq}{\begin{equation}}
\newcommand{\eeq}[1]{\label{#1}\end{equation}}
\newcommand{\ber}{\begin{eqnarray}}
\newcommand{\eer}[1]{\label{#1}\end{eqnarray}}
\begin{titlepage}
\begin{center}

\hfill IMAFF-FM-97/03\\
\hfill NIKHEF-97-033\\
\hfill hep-th/9707211\\
\vskip .5in

{\large \bf The Even and the Odd Spectral Flows on the N=2 Superconformal
 Algebras}
\vskip .8in

{\bf Beatriz Gato-Rivera}\\

\vskip .3in

{\em Instituto de Matem\'aticas y F\'\i sica Fundamental, CSIC,\\ Serrano 123,
Madrid 28006, Spain} \footnote{e-mail addresses:
bgato@pinar1.csic.es}\\

\vskip .3in

{\em NIKHEF, Kruislaan 409, NL-1098 SJ Amsterdam, The Netherlands}\\

\end{center}

\vskip .6in

\begin{center} {\bf ABSTRACT } \end{center}
\begin{quotation}
There are two different spectral flows on the N=2 superconformal algebras
(four in the case of the Topological algebra). The usual spectral flow, first 
considered by Schwimmer and Seiberg, is an even transformation, whereas the
spectral flow previously considered by the author and Rosado is an odd
transformation. We show that the even spectral flow is generated by the odd
spectral flow, and therefore only the latter is fundamental. We also analyze
thoroughly the four ``topological'' spectral flows, writing two of them here
for the first time. Whereas the even and the odd spectral flows have
quasi-mirrored properties acting on the Antiperiodic or the Periodic algebras,
the topological even and odd spectral flows have drastically different
properties acting on the Topological algebra. The other two topological
spectral flows have mixed even and odd properties. We show that the even
and the even-odd topological spectral flows are generated by the odd and the
odd-even topological spectral flows, and therefore only the latter are
fundamental.

\end{quotation}
\vskip .6cm
July 1997

\end{titlepage}
\vfill
\eject
\def\baselinestretch{1.2}
\baselineskip 17 pt
\section{Introduction}\lvm

The N=2 Superconformal algebras provide the symmetries underlying the
N=2 strings \cite{Adem}.
In addition, the topological version of the algebra is realized in the
world-sheet of the bosonic string \cite{BeSe}, as well as in the world-sheet
of the superstrings \cite{BLNW}. The spectral flows, which are
denoted as $\cU_{\th}$ and $\cA_{\th}$, transform the N=2
Superconformal algebras into isomorphic algebras for any value of $\th$.
In particular, for half-integer values of $\th$ they interchange
the Antiperiodic NS algebra
and the Periodic R algebra. For specific values of this parameter they
also transform primary states into primary states (and singular vectors
into singular vectors, consequently). As we will show, the spectral flow
$\cU_{\th}$ (the usual spectral flow)
is an even transformation while the spectral flow $\cA_{\th}$ is odd.

The even spectral flow $\cU_{\th}$ was written down in \cite{SS}. It has
been considered in various papers, for example
 \cite{Kir1}\cite{LVW}\cite{Bert}\cite{BJI4}. The
topological twisting of this spectral flow was analyzed for the first
time in \cite{BJI3}, although only the applications to the chiral
representations of the Topological algebra were taken into account. Some
other applications of the ``topological even spectral flow" can be seen in
\cite{BJI6} and \cite{SeTi}, although an appropriate analysis of the issue
has never been done.

The odd spectral flow $\cA_{\th}$
was written down in \cite{BJI4}, where it was denoted as ``the other" spectral
flow. The topological twisting of this spectral flow has never been
considered, although the ``topological odd spectral flow" for $\th=1$, which
is an automorphism of the Topological algebra denoted simply as $\cA$, has
been applied in several occasions to map topological singular vectors into
each other \cite{BJI3}\cite{BJI6}\cite{BJI7}.

In this paper we intend to fill the existing gaps completing the whole
picture. In section 2 we first review the main properties of the
even and the odd spectral flows acting on the Antiperiodic NS and the
Periodic R algebras, giving some new insights. Then we present the
composition rules of the two spectral flows, showing that the even
spectral flow $\cU_{\th}$ is generated by the odd spectral flow
$\cA_{\th}$, which is therefore the only fundamental one.
In section 3 we analyze thoroughly the topological twistings
of the two spectral flows, giving rise to four different ``topological''
spectral flows, denoted as $\cU_{\th}$, $\cA_{\th}$, $\hat\cU_{\th}$ and
$\hat\cA_{\th}$, two of which are written here for the first time. We
point out the main properties of the four topological spectral flows and we
write down their composition rules. We find that $\cU_{\th}$ is even,
$\cA_{\th}$ is odd, $\hat\cU_{\th}$ has even-odd properties, and
$\hat\cA_{\th}$ has odd-even properties.
The odd and the odd-even spectral flows generate the even and the
even-odd spectral flows, and therefore they
are the only fundamental topological spectral flows.
 In section 4 we write the conclusions
and final remarks.

\section{Spectral Flows on the N=2 {\ }NS and R Algebras}\lvm

The N=2 Superconformal algebra
\cite{Adem}\cite{DiVPZ}\cite{Kir1}\cite{LVW} can be expressed as

\BE\new\BA{lclclcl}
\L[L_m,L_n\R]&=&(m-n)L_{m+n}+{\ctop\over12}(m^3-m)\Kr{m+n}
\,,&\qquad&[H_m,H_n]&=
&{\ctop\over3}m\Kr{m+n}\,,\\
\L[L_m,G_r^\pm
\R]&=&\L({m\over2}-r\R)G_{m+r}^\pm
\,,&\qquad&[H_m,G_r^\pm]&=&\pm G_{m+r}^\pm\,,
\\
\L[L_m,H_n\R]&=&{}-nH_{m+n}\\
\L\{G_r^-,G_s^+\R\}&=&\multicolumn{5}{l}{2L_{r+s}-(r-s)H_{r+s}+
{\ctop\over3}(r^2-\frac{1}{4})
\Kr{r+s}\,,}\EA\label{N2alg}
\EE

\noi
where $L_m$ and $H_m$ are the spin-2 and spin-1 bosonic generators
corresponding to the stress-energy momentum tensor and the U(1) current,
respectively, and
$G_r^+$ and $G_r^-$ are the spin-3/2 fermionic generators. These
are half-integer moded for the case of the Antiperiodic NS algebra, and
integer moded for the case of the Periodic R algebra.

In order to simplify the analysis that follows we will
unify the notation for the $U(1)$
charge of the states of the Antiperiodic NS
algebra and the states of the Periodic R algebra.
Namely, the $U(1)$ charge of the Ramond states will be
denoted by $h$, instead of $h\pm \half$, like the U(1) charge of the
NS states. In addition, the
relative charge $q$ of a secondary state will be defined as
the difference between the $U(1)$ charge of the state and the
$U(1)$ charge of the primary on which it is built.
Therefore, the relative charges of the Ramond states are
defined to be integer, like the relative charges of the NS states.

\subsection{The Even and the Odd Spectral Flows}\lvm

The spectral flows $\,\cU_{\th}$ and $\cA_{\th}$ are
one-parameter families of transformations providing a continuum of
isomorphic N=2 Superconformal algebras. The ``usual'' spectral flow
$\cU_{\th}$ \cite{SS}\cite{LVW}\cite{BJI4} is even, given by

\BE\new\BA{rclcrcl}
\cU_\th \, L_m \, \cU_\th^{-1}&=& L_m
 +\th H_m + {\ctop\over 6} \th^2 \delta_{m,0}\,,\\
\cU_\th H_m \, \cU_\th^{-1}&=&H_m + {\ctop\over3} \th \delta_{m,0}\,,\\
\cU_\th \, G^+_r \, \cU_\th^{-1}&=&G_{r+\th}^+\,,\\
\cU_\th \, G^-_r \, \cU_\th^{-1}&=&G_{r-\th}^-\,,\
\label{spfl} \EA\EE

\noi
satisfying $\cU^{-1}_\th = \cU_{(-\th)}$. For $\th=0$ it is just the
identity operator, \ie\ $\cU_0={\bf 1}$. It transforms the $(L_0, H_0)$
eigenvalues, \ie\ the conformal weight and the
U(1) charge, $(\Delta, h)$ of a given state as
 $(\Delta-\th h +{\ctop\over6} \th^2, \,h- {\ctop\over3} \th)$. From this
one gets straightforwardly that the level $l$ of any secondary state
changes to $l-\th q$, while the relative charge $q$ remains equal.

The spectral flow $\cA_{\th}$ \cite{BJI4} is odd, given by

\BE\new\BA{rclcrcl}
\cA_\th \, L_m \, \cA_\th^{-1}&=& L_m
 +\th H_m + {\ctop\over 6} \th^2 \delta_{m,0}\,,\\
\cA_\th H_m \, \cA_\th^{-1}&=&- H_m - {\ctop\over3} \th \delta_{m,0}\,,\\
\cA_\th \, G^+_r \, \cA_\th^{-1}&=&G_{r-\th}^-\,,\\
\cA_\th \, G^-_r \, \cA_\th^{-1}&=&G_{r+\th}^+\,,\
\label{ospfl} \EA\EE

\noi
satisfying $\cA_{\th}^{-1} = \cA_{\th}$. It is therefore an involution.
$\cA_{\th}$ is ``quasi'' mirror symmetric to $\cU_{\th}$:
under the exchange $H_m \to -H_m$, $ G_r^+ \leftrightarrow G_r^-$ and
$\th \to -\th$. For $\th=0$ it is the mirror map, \ie\ $\cA_0={\cal M}$.
It transforms the $(L_0, H_0)$ eigenvalues of the states as
$(\Delta+\th h +{\ctop\over6} \th^2, - h - {\ctop\over3} \th)$.
The level $l$ of the secondary states changes
to $l + \th q$, while the relative charge $q$ reverses its sign.

\vskip .2in

The table below summarizes the properties of $\cU_{\th}$ and $\cA_{\th}$
for general values of $\th$. Observe that $\cU_{\th}\, \kc^{(q)}_l$
and $\cA_{\th}\, \kc^{(q)}_l$ are not mirror symmetric because of the sign
of $\th$.

\vskip .5in

\begin{tabular}{r|c c c r}
{\ } & conformal weight & U(1) charge & level & relative charge \\
\hline \\
$\kc^{(q)}_l$ & $\D$ & $h$ & $l$ & $q\ \ \ \ \ \ \ $ \\
$\cU_{\th} \, \kc^{(q)}_l$ & $\D-\th h+{\ctop\over6}\th^2$ &
$h- {\ctop\over3} \th$ & $l-\th q$ & $q\ \ \ \ \ \ \ $ \\
$\cA_{\th} \, \kc^{(q)}_l$ & $\D+\th h+{\ctop\over6}\th^2$ &
$- h - {\ctop\over3} \th \ \ $ & $l+\th q$ & $-q\ \ \ \ \ \ \ $\\
 \end{tabular}

\vskip .6in

\subsection{Main Properties}

Let us review the main properties of the spectral flows $\cU_{\th}$ and
$\cA_{\th}$ for the most interesting values of $\th$, those mapping
primary states to primary states, adding some new insights.
More details can be found in refs. \cite{SS}, \cite{LVW}, \cite{BJI4} and
\cite{BJI7}.

\vskip .16in
For half-integer values of $\th$ the two spectral flows interpolate between
the Antiperiodic NS algebra and the Periodic R algebra. In particular, for
$\th = \pm 1/2$ the primary states of the NS algebra
are transformed into primary states of the R
algebra with helicities $(\mp)$(\ie\ annihilated by $G^-_0$ and $G^+_0$,
respectively). As a result the NS singular vectors are transformed into R
singular vectors with helicities $(\mp)$ built on R primaries with the
same helicities. In addition, $\cU_{1/2}$ and $\cA_{-1/2}$ map the chiral
NS primaries (annihilated by $G^+_{-1/2}$) to the set of
R ground states (annihilated by both $G^+_0$ and $G^-_0$),
whereas $\cU_{-1/2}$ and $\cA_{1/2}$ map the antichiral NS
primaries (annihilated by $G^-_{-1/2}$) to the R ground states. As a
consequence $\cU_{1/2}$ and $\cA_{-1/2}$ transform the NS singular vectors
built on chiral primaries into helicity $(\mp)$ R singular vectors
built on R ground states, respectively,
whereas $\cU_{-1/2}$ and $\cA_{1/2}$ transform the NS singular vectors
built on antichiral primaries into helicity $(\pm)$ R singular vectors
built on R ground states, respectively.

\vskip .16in
For integer values of $\th$ the NS algebra and the R algebra map back to
themselves, although the states that were primary with respect to the
initial algebra are in general not primary with respect to the final algebra. 

In the case of the NS algebra only primary states that
are chiral or antichiral can be mapped back into NS primary states, which
turn out to be also chiral or antichiral.
In addition, only $\cU_1$ and
$\cA_0$ map chiral primaries into antichiral primaries, only
$\cU_{-1}$ and $\cA_0$ map antichiral primaries into chiral primaries,
only $\cA_{-1}$ and $\cU_0$ map chiral primaries
into chiral primaries, and only $\cA_1$ and $\cU_0$ map
antichiral primaries into antichiral primaries. An interesting
consequence of this is that for no value of
$\th$, except $\th=0$, the spectral flows map
singular vectors of the NS algebra back into NS singular vectors. In other
words, the identity and the mirror map are the only spectral flows that
transform NS singular vectors back into NS singular vectors. This is
due to the fact that there are no chiral singular vectors built on chiral
primaries neither antichiral singular vectors built on antichiral primaries.
Observe that for no values of $\th\neq0$
does the odd spectral flow $\cA_{\th}$
interpolate between the chiral ring and the antichiral ring, and inversely,
for no values of $\th\neq0$ does the even spectral flow $\cU_{\th}$ map
the chiral and the antichiral rings back to themselves.

In the case of the R algebra only $\cU_{\pm1}$ and $\cA_0$
transform primary states with helicity $(\pm)$
into primary states with helicity $(\mp)$, whereas only
$\cA_{\pm1}$ and $\cU_0$ transform primary states with helicity $(\mp)$
back into helicity $(\mp)$ primaries (observe that $\cA_{\pm1}$ does
not reverse the helicity as $\cU_{\pm1}$ does).
As a consequence, all the singular vectors of the R algebra with the same
helicity as the primaries on which they are built (and only these)
can be mapped back into R singular vectors using spectral flows with either
$\th=1$ or $\th=-1$,
in addition to the identity $\cU_0$ and the mirror map $\cA_0$, which
transform all kinds of R singular vectors into R singular vectors.
Regarding the R ground states, for no value of $\th\neq0$ they are
mapped back to R ground states. Under $\cU_{\pm1}$ and $\cA_{\pm1}$
they are transformed into helicity $(\mp)$ primaries with the additional
condition of being annihilated by $G^{\mp}_{-1}$.

\subsection{Composition Rules}\lvm

The composition rules are the following. For the even spectral flow
one has simply

\BE  \cU_{\th_2} \ \cU_{\th_1} = \cU_{(\th_2 + \th_1)} , \label{crev}\EE

\noi
from which one obtains ${\ }\cU_0 = \o1{\ }$ and
${\ }\cU_{\th}^{-1} = \cU_{(-\th)}$. For the odd spectral flow one finds

\BE  \cA_{\th_2} \ \cA_{\th_1} = \cU_{(\th_2 - \th_1)} ,  \EE

\BE  \cA_{\th_2} \ \cU_{\th_1} = \cA_{(\th_2 - \th_1)} , \qquad
\cU_{\th_2} \ \cA_{\th_1} = \cA_{(\th_2 + \th_1)} ,  \EE

\noi
from which one obtains ${\ }\cA_{\th}^{-1} = \cA_{\th}{\ }$, as well as the
relations

\BE \cU_{\th} = \cA_{\th} \ \cA_0 \ = \cA_0 \ \cA_{(-\th)} \ ,  \EE

\BE \cA_{\th} = \cU_{\th} \ \cA_0 \ = \cA_0 \ \cU_{(-\th)} \ ,  \qquad
\cA_0 = \cA_{\th} \ \cU_{\th} = \cU_{\th} \ \cA_{(-\th)} \ . \EE

\noi
$\cA_0$  is the mirror map $H_m \leftrightarrow -H_m$,
$G_r^+ \leftrightarrow G_r^-$, as we pointed out before.
We see that the odd spectral flow $\cA_{\th}$ generates the even spectral
flow $\,\cU_{\th}$, and therefore it is the only fundamental spectral flow.
Observe that $\cU_{\th}$ and $\cA_{\th}$ do not commute, and $\cU_{\th}$
commutes with itself whereas $\cA_{\th}$ does not.
Notice also that $\cU_{\th}^{-1} = \cU_{(-\th)}$ while
$\cA_{\th}^{-1} = \cA_{\th}$, \ie\ the inverse of the even
spectral flow with parameter $\th$ is the one with parameter $(-\th)$,
while the odd spectral flow is its own inverse, \ie\ an involution.

\section{Spectral Flows on the N=2 Topological Algebra}\lvm

The Topological N=2 Superconformal algebra reads \cite{DVV}

\BE\new\BA{lclclcl}
\L[\cL_m,\cL_n\R]&=&(m-n)\cL_{m+n}\,,&\qquad&[\cH_m,\cH_n]&=
&{\ctop\over3}m\Kr{m+n}\,,\\
\L[\cL_m,\cG_n\R]&=&(m-n)\cG_{m+n}\,,&\qquad&[\cH_m,\cG_n]&=&\cG_{m+n}\,,
\\
\L[\cL_m,\cQ_n\R]&=&-n\cQ_{m+n}\,,&\qquad&[\cH_m,\cQ_n]&=&-\cQ_{m+n}\,,\\
\L[\cL_m,\cH_n\R]&=&\multicolumn{5}{l}{-n\cH_{m+n}+{\ctop\over6}(m^2+m)
\Kr{m+n}\,,}\\
\L\{\cG_m,\cQ_n\R\}&=&\multicolumn{5}{l}{2\cL_{m+n}-2n\cH_{m+n}+
{\ctop\over3}(m^2+m)\Kr{m+n}\,,}\EA\qquad m,~n\in\oZ\,.\label{topalgebra}
\EE

\noi
where the fermionic generators $\cQ_m$ and $\cG_m$ correspond
to the spin-1 BRST current and the spin-2 fermionic current, respectively.
The Topological algebra \req{topalgebra} can be viewed as a rewriting of
the algebra \req{N2alg} using one of the two topological twists:

\BE\new\BA{rclcrcl}
\cL^{(1)}_m&=&\multicolumn{5}{l}{L_m+\half(m+1)H_m\,,}\\
\cH^{(1)}_m&=&H_m\,,&{}&{}&{}&{}\\
\cG^{(1)}_m&=&G_{m+\half}^+\,,&\qquad &\cQ_m^{(1)}&=&G^-_{m-\half}
\,,\label{twa}\EA\EE

\noi
and

\BE\new\BA{rclcrcl}
\cL^{(2)}_m&=&\multicolumn{5}{l}{L_m-\half(m+1)H_m\,,}\\
\cH^{(2)}_m&=&-H_m\,,&{}&{}&{}&{}\\
\cG^{(2)}_m&=&G_{m+\half}^-\,,&\qquad &
\cQ_m^{(2)}&=&G^+_{m-\half}\,,\label{twb}\EA\EE

\noi
which we denote as $T_{W1}$ and $T_{W2}$, respectively. Observe that the
two twists are mirrored. In particular $(G^{+}_{1/2}, G^{-}_{-1/2})$
results in $(\cG^{(1)}_0, \cQ^{(1)}_0)$, while
$(G^{-}_{1/2}, G^{+}_{-1/2})$ gives $(\cG^{(2)}_0, \cQ^{(2)}_0)$,
so that the topological chiral primaries (annihilated by both $\cQ_0$
and $\cG_0$) correspond to the antichiral and the chiral primaries of the
NS algebra under the twists $T_{W1}$ and $T_{W2}$, respectively.

\subsection{The Topological Spectral Flows}\lvm

The ``topological'' spectral flows are obtained by twisting the spectral
flows \req{spfl} and \req{ospfl}. There are two ways to proceed in each case:
either using the same twist on the left and on the right-hand sides of
expressions \req{spfl} and \req{ospfl}, or using different twists on the
left and on the right. In the first case one obtains the even and the odd
spectral flows

\BE\new\BA{rclcrcl}
\cU_\th \, \cL^{(1)}_m \, \cU_\th^{-1}&=& \cL^{(1)}_m
 +\th \cH^{(1)}_m + {\ctop\over6}(\th^2+\th) \delta_{m,0}\,,\\
\cU_{\th} \, \cH^{(1)}_m \, \cU_\th^{-1}&=&\cH^{(1)}_m +
 {\ctop\over3} \th \delta_{m,0}\,,\\
\cU_\th \, \, \cG^{(1)}_m \, \cU_\th^{-1}&=&\cG^{(1)}_{m+\th}\,,\\
\cU_\th \, \cQ^{(1)}_m \, \cU_\th^{-1}&=&\cQ^{(1)}_{m-\th}\,,\\
\label{tspfl} \EA\EE

\noi
with $\cU_{\th}^{-1} = \cU_{(-\th)}$, and

\BE\new\BA{rclcrcl}
\cA_\th \, \cL^{(1)}_m \, \cA_\th^{-1}&=& \cL^{(1)}_m
 +(\th - m-1)\cH^{(1)}_m + {\ctop\over6}(\th^2-\th) \delta_{m,0}\,,\\
\cA_\th \, \cH^{(1)}_m \, \cA_\th^{-1}&=&-\cH^{(1)}_m -
 {\ctop\over3} \th \delta_{m,0}\,,\\
\cA_\th \, \, \cG^{(1)}_m \, \cA_\th^{-1}&=&\cQ^{(1)}_{m+1-\th}\,,\\
\cA_\th \, \cQ^{(1)}_m \, \cA_\th^{-1}&=&\cG^{(1)}_{m-1+\th}\,,\\
\label{tospfl} \EA\EE

\noi
with $\cA_{\th}^{-1} = \cA_{\th}$.
One obtains the same expressions for the generators with label (2)
but with $\th \to -\th$. These topological spectral flows satisfy
the same composition rules as the even and the odd untwisted spectral flows
\req{spfl} and \req{ospfl}. For this reason we denote them in the same way.

The topological even spectral flow $\cU_\th$ \cite{SeTi}\cite{BJI6}
looks (and behaves) almost
identical as its untwisted partner \req{spfl}. It transforms the
 $(\cL_0, \cH_0)$ eigenvalues $(\Delta, h)$ of any given state as
 $(\Delta-\th h +{\ctop\over6} (\th^2-\th), \, h- {\ctop\over3} \th)$.
The level $l$ of the secondary states changes
to $l-\th q$, whereas the relative charge $q$ remains unchanged.

The topological odd spectral flow $\cA_\th$ looks and behaves quite
differently from its untwisted partner \req{ospfl}. It has never been
considered in the literature before, although the specific transformation
$\cA_1$, denoted simply as $\cA$, has
been used in several papers \cite{BJI3}\cite{BJI6}\cite{BJI7} to map
singular vectors into each other. $\cA_\th$ transforms the
 $(\cL_0, \cH_0)$ eigenvalues $(\Delta, h)$ of a given state as
 $(\Delta + (\th-1)h +{\ctop\over6} (\th^2-\th), -h-{\ctop\over3} \th)$.
The level $l$ of the secondary states changes
to $l + (\th-1) q$, and the relative charge $q$ reverses its sign.

\vskip .17in
By using different twists on the left and on the right-hand sides of the
spectral flows \req{spfl} and \req{ospfl} one obtains the expressions

\BE\new\BA{rclcrcl}
\hat\cU_\th \, \cL^{(1)}_m \, \hat\cU_\th^{-1}&=& \cL_m^{(2)}
 -(\th + m+1)\cH_m^{(2)} + {\ctop\over6}(\th^2+\th) \delta_{m,0}\,,\\
\hat\cU_\th \, \cH^{(1)}_m \, \hat\cU_\th^{-1}&=&-\cH_m^{(2)} +
 {\ctop\over3} \th \delta_{m,0}\,,\\
\hat\cU_\th \, \, \cG^{(1)}_m \, \hat\cU_\th^{-1}&=&\cQ_{m+1+\th}^{(2)}\,,\\
\hat\cU_\th \, \cQ^{(1)}_m \, \hat\cU_\th^{-1}&=&\cG_{m-1-\th}^{(2)}\,,\\
\label{tspfl12} \EA\EE

\noi
and

\BE\new\BA{rclcrcl}
\hat\cA_\th \, \cL^{(1)}_m \, \hat\cA_\th^{-1}&=& \cL_m^{(2)}
 -\th \cH_m^{(2)} + {\ctop\over6}(\th^2-\th) \delta_{m,0}\,,\\
\hat\cA_\th \, \cH^{(1)}_m \, \hat\cA_\th^{-1}&=&\cH_m^{(2)} -
 {\ctop\over3} \th \delta_{m,0}\,,\\
\hat\cA_\th \, \, \cG^{(1)}_m \, \hat\cA_\th^{-1}&=&\cG_{m-\th}^{(2)}\,,\\
\hat\cA_\th \, \cQ^{(1)}_m \, \hat\cA_\th^{-1}&=&\cQ_{m+\th}^{(2)}\,.\\
\label{tospfl12} \EA\EE

\noi
One exchanges the labels $(1) \leftrightarrow (2)$ in these
expressions by setting $\th \to -\th$ simultaneously. The spectral flows
$\hat\cU_{\th}$ and $\hat\cA_{\th}$, which we denote as even-odd and
odd-even respectively, connect the generators of the topological theory
(1) to the generators of the topological theory (2),
having mixed even and odd properties.
Their inverses, which connect the generators of
the topological theory (2) to the generators of the topological theory (1),
are given by $\hat\cU_{\th}^{-1} = \hat\cU_{(-\th)}$ and
$\hat\cA_{\th}^{-1} = \hat\cA_{\th}$, \ie\ the same expressions as for
the even and the odd spectral flows, respectively.

The topological even-odd spectral flow $\hat\cU_\th$ looks
and behaves very similarly as the topological odd spectral flow $\cA_\th$
\req{tospfl}, in spite of not being an involution. It was written down in
ref. \cite{BJI3}, although only the values $\th=\pm1$ were taken into
account\footnote{In ref. \cite{BJI3} only topological chiral primaries
were considered. They are mapped into each other by $\hat\cU_\th$ only
for $\th=\pm1$.}.
It transforms the  $(\cL_0^{(1)}, \cH_0^{(1)})$ eigenvalues
$(\Delta^{(1)}, h^{(1)})$ of the states of the topological theory (1)
into $(\cL_0^{(2)}, \cH_0^{(2)})$ eigenvalues of the states of the
topological theory (2) as $(\Delta^{(2)}, h^{(2)}) =
(\Delta^{(1)}-(\th+1) h^{(1)}+{\ctop\over6}(\th^2+\th),
-h^{(1)}+{\ctop\over3}\th)$. It modifies the level $l$ of the secondary
states as $l^{(2)}=l^{(1)}-(\th+1)q^{(1)}$, reversing
the sign of the relative charge, \ie\ $q^{(2)}=-q^{(1)}$.

The topological odd-even spectral flow $\hat\cA_\th$ looks
and behaves very similarly as the even spectral flows $\cU_\th$, in spite
of being an involution. It has never been considered in the literature
before. It transforms the  $(\cL_0^{(1)}, \cH_0^{(1)})$ eigenvalues
$(\Delta^{(1)}, h^{(1)})$ of the states into $(\cL_0^{(2)}, \cH_0^{(2)})$
eigenvalues as $(\Delta^{(2)}, h^{(2)}) =
(\Delta^{(1)} + \th h^{(1)} +{\ctop\over6}(\th^2+\th),\,
 h^{(1)} + {\ctop\over3} \th)$.
It modifies the level $l$ of the secondary states as
$l^{(2)}=l^{(1)}+\th q^{(1)}$, letting the relative charge invariant, \ie\
$q^{(2)}=q^{(1)}$.

\vskip .2in

The table below summarizes the properties of the topological spectral
flows $\,\cU_{\th}$, $\cA_{\th}$, $\,\hat\cU_{\th}$ and $\hat\cA_{\th}$,
for general values of $\th$, acting on the states of the topological
theory (1). One finds the same table for the spectral flows acting on the
states of the topological theory (2) but with $\th \to -\th$.

\vskip .5in

\begin{tabular}{r|c c c r}
{\ } & conformal weight & U(1) charge & level & relative charge \\
\hline \\
$\kca^{(q)}_l$ & $\D$ & $h$ & $l$ & $q\ \ \ \ \ \ \ $ \\
$\cU_{\th}\,\kca^{(q)}_l$ &
$\D-\th h +{\ctop\over6} (\th^2-\th)$ &
$h- {\ctop\over3} \th$ & $l-\th q$ & $q\ \ \ \ \ \ \ $ \\
$\cA_{\th}\,\kca^{(q)}_l$ &
$\D + (\th-1)h +{\ctop\over6} (\th^2-\th)$ &
$- h - {\ctop\over3} \th \ \ $ & $l+ (\th-1) q$ & $-q\ \ \ \ \ \ \ $ \\
$\hat\cU_{\th}\,\kca^{(q)}_l$ & $\D-(\th+1) h+{\ctop\over6}(\th^2+\th) $ &
$-h+{\ctop\over3}\th \ \ $ & $l-(\th+1)q$ & $-q\ \ \ \ \ \ \ $ \\
$\hat\cA_{\th}\,\kca^{(q)}_l$ & $\D+\th h+{\ctop\over6}(\th^2+\th)$ &
$h+{\ctop\over3} \th$ &  $l+\th q$ & $q\ \ \ \ \ \ \ $ \\
 \end{tabular}

\vskip .6in

\subsection{Main Properties}\lvm

Now let us discuss the main properties of the topological spectral
flows $\,\cU_{\th}$, $\cA_{\th}$, $\,\hat\cU_{\th}$ and $\hat\cA_{\th}$
for the most interesting values of $\th$, those mapping primary states to
primary states.

\vskip .25in
{\it The Even Spectral Flow} $\cU_\th$

\vskip .12in
The action of the topological even spectral flow $\cU_\th$ \req{tspfl} on
the states of the Topological algebra is very similar to the action of its
untwisted partner $\cU_\th$ \req{spfl} on the states of the Periodic R
algebra. Apart from the identity $\cU_0$, only $\cU_{\pm1}$
map primary states into primary states, under restricted conditions though.
Under $\cU_1$ only \Gn-closed primaries (\ie\ annihilated by \Gn)
are transformed back into primaries. These turn out to be \Qn-closed
(\ie\ annihilated by \Qn\ and therefore BRST-invariant). Inversely, under
$\cU_{-1}$ only \Qn-closed primaries are transformed into primaries, which
turn out to be \Gn-closed. As a consequence, under $\cU_1$ only \Gn-closed
singular vectors built on \Gn-closed primaries are transformed into
singular vectors as well. These turn out to be \Qn-closed built on \Qn-closed
primaries. The inverse situation occurs under $\cU_{-1}$.

For no value of $\th\neq0$ do the topological chiral primaries
(\ie\ annihilated by \Gn\ and \Qn) transform back into topological chiral
primaries. Under $\cU_1$ they transform into \Qn-closed primaries with the
additional constraint of being annihilated by $\cQ_{-1}$, whereas under
$\cU_{-1}$ they transform into \Gn-closed primaries with the additional
constraint of being annihilated by $\cG_{-1}$. Hence the topological
even spectral flow $\cU_\th$ ``destroys'' topological chiral Verma
modules\footnote{Curiously, this spectral flow has been considered in
ref. \cite{SeTi} acting on topological chiral Verma modules.}.
Chiral singular vectors in turn, built on \Gn-closed or \Qn-closed
primaries (they do not exist on chiral primaries) transform under $\cU_1$
or $\cU_{-1}$, respectively, into non-chiral singular vectors
annihilated by $\cQ_{-1}$ or $\cG_{-1}$, as a result.

\vskip .25in
{\it The Odd Spectral Flow} $\cA_\th$

\vskip .12in
The action of the topological odd spectral flow $\cA_\th$ \req{tospfl} on
the states of the Topological algebra is drastically different from the
action of the topological even spectral flow $\cU_\th$ and also drastically
different from the action of the untwisted odd spectral flow $\cA_\th$
\req{ospfl} on the states of the Periodic R algebra and the states of the
Antiperiodic NS algebra. The main difference consists of the existence of
a value of $\th$, namely $\th=1$, for which the topological odd spectral flow
becomes a ``universal'' transformation, in the sense that all kinds of
primary states are mapped back into primary states. In addition, the
Topological algebra automorphism $\cA$, as $\cA_1$ is denoted
\cite{BJI3}\cite{BJI6}\cite{BJI7}, transforms topological chiral primaries
into topological chiral primaries. As a consequence, $\cA$ transforms
all kinds of singular vectors into singular vectors, and singular vectors
in topological chiral Verma modules back to singular vectors in
topological chiral Verma modules.

Let us say a few more words about the Topological algebra automorphism
$\cA$, because of its importance. It reads \cite{BJI3}

\BE\new\BA{rclcrcl}
\cA \, \cL_m \, \cA&=& \cL_m - m\cH_m\,,\\
\cA \, \cH_m \, \cA&=&-\cH_m - {\ctop\over3} \delta_{m,0}\,,\\
\cA \, \,\cG_m \, \cA&=&\cQ_m\,,\\
\cA \, \cQ_m \, \cA&=&\cG_m\,.\

\label{autom} \EA\EE

\noi
It transforms the $(\cL_0, \cH_0)$ eigenvalues $(\Delta, h)$ of the states
as $(\Delta, - h - {\ctop\over3})$. Hence it does not modify the conformal
weight. As a result the level of the secondary states remains unchanged
whereas the relative charge is reversed ($q \to -q$). In addition $\cA$
also reverses the BRST-invariance properties of the states: \Gn-closed
states are mapped to \Qn-closed states, and the other way around, and
chiral states are mapped to chiral states, consequently.

There are two other values of $\th$ for which $\cA_\th$ maps primary states
into primary states although with restrictions: $\th=0$ and $\th=2$.
$\cA_0$ transforms only \Gn-closed primary states back into primary states,
which are \Gn-closed as well\footnote{$\cA_0$ has been considered in  ref.
\cite{RS} where it was denoted as the topological mirror map. However the
authors treated $\cA_0$ as an isolated transformation, not as a particular
case of a family of transformations, \ie\ of a spectral flow.},
mapping chiral primary states to non-chiral \Gn-closed primaries
annihilated by $\cG_{-1}$. The complementary mapping is performed
by $\cA_2\,$; that is, it transforms only \Qn-closed primary states into
primary states, which are \Qn-closed as well, mapping chiral primaries
to non-chiral \Qn-closed primaries annihilated by $\cQ_{-1}$.
As a consequence only \Gn-closed (\Qn-closed) singular vectors built on
\Gn-closed (\Qn-closed) primaries are transformed by $\cA_0$ ($\cA_2$)
back into singular vectors. These turn out to be again \Gn-closed
(\Qn-closed) singular vectors built on \Gn-closed (\Qn-closed) primaries.
Chiral singular vectors are transformed in the same manner, with
the additional constraint of being annihilated by $\cG_{-1}$ ($\cQ_{-1}$).

Observe the striking differences between the topological $\cA_0$
and the untwisted $\cA_0$, which is the mirror map transforming all
the primaries and singular vectors of the NS algebra and the R algebra
into mirrored primaries and mirrored singular vectors, and the chiral
and the antichiral primaries into each other.

\vskip .25in
{\it The Even-Odd Spectral Flow} $\hat\cU_\th$

\vskip .12in
The topological spectral flow $\hat\cU_\th$ \req{tspfl12} is very
similar to the odd spectral flow $\cA_\th$ \req{tospfl}, although it is
not an involution and satisfy the same composition rules as the
even spectral flows. For this reason we denote it as ``even-odd''.
For $\th=0$ it is the ``identity'' which gives the
exact relation between the generators of the topological theory (1) and
the generators of the topological theory (2), as deduced from the twists
\req{twa} and \req{twb}. However, $\hat\cU_0$ transforms only the \Gn-closed
primary states of one theory into primary states of the other theory,
which are also \Gn-closed, in complete analogy with the action of $\cA_0$.
The complementary transformations mapping \Qn-closed primary states into
\Qn-closed primary states are given by $\hat\cU_{\pm2}$ ($\th=2$ for the
mapping from the states of theory (2) to the states of theory (1), and the
other way around for $\th=-2$). As a consequence, only \Gn-closed singular
vectors built on \Gn-closed primaries are transformed into singular vectors
under $\hat\cU_0$, which are also \Gn-closed built on \Gn-closed primaries.
Similarly, only \Qn-closed singular vectors built on \Qn-closed primaries
are transformed under $\hat\cU_{\pm2}$ into singular vectors, which are
also \Qn-closed built on \Qn-closed primaries. Chiral primaries, and chiral
singular vectors built on \Gn-closed or \Qn-closed primaries, are
transformed under $\hat\cU_0$ and $\hat\cU_{\pm2}$ into non-chiral
\Gn-closed and \Qn-closed primaries, and singular vectors, annihilated
by $\cG_{-1}$ and $\cQ_{-1}$, respectively.

In turn, $\hat\cU_{\pm1}$ are the ``universal'' mappings which transform
all kinds of primary states of one theory into primary states of the other
theory, and chiral primary states into chiral primary states ($\th=1$
for the mapping from the states of theory (2) to the states of theory (1),
and the other way around for $\th=-1$). As a result all the singular vectors
of each theory are mapped into singular vectors of the other theory.
In particular singular vectors in chiral Verma modules of one theory are
transformed into singular vectors in chiral Verma modules of the other
theory. Let us write $\hat\cU_1$ explicitely, because of its importance:

\BE\new\BA{rclcrcl}

\hat\cU_1 \, \cL^{(2)}_m \, \hat\cU_1^{-1}&=& \cL_m^{(1)} - m\cH_m^{(1)}\,,\\
\hat\cU_1 \, \cH^{(2)}_m \, \hat\cU_1^{-1}&=&-\cH_m^{(1)}
 -{\ctop\over3} \delta_{m,0}\,,\\
\hat\cU_1 \, \,\cG^{(2)}_m \, \hat\cU_1^{-1}&=&\cQ_m^{(1)}\,,\\
\hat\cU_1 \, \cQ^{(2)}_m \, \hat\cU_1^{-1}&=&\cG_m^{(1)}\,.\
\label{spfla}\EA\EE

\noi
We see that $\hat\cU_1$, with the generators of the topological theory (2)
on the left-hand side, looks exactly like the Topological algebra
automorphism\footnote{As a matter of fact, the Topological algebra
automorphism $\cA$ was found in ref. \cite{BJI3} just by erasing
the labels (1) and (2) in expression \req{spfla}.} $\cA$ \req{autom}.

\vskip .25in
{\it The Odd-Even Spectral Flow} $\hat\cA_\th$

\vskip .12in
The topological spectral flow $\hat\cA_\th$ \req{tospfl12} is very
similar to the even spectral flow $\cU_\th$ \req{tspfl}, although it is
an involution and satisfy the same composition rules as the odd spectral
flows. For this reason we denote it as ``odd-even''.
For $\th=0$ it just produces the
interchange of labels $(1) \leftrightarrow (2)$;
that is, the interchange between the two topological
theories. Therefore $\hat\cA_0$ maps primary states and singular vectors
between the two topological theories in a trivial way. For $\th=\pm1$ it
transforms primaries of one theory into primaries of the other theory under
restricted conditions: $\hat\cA_1$ maps $\Qz^{(1)}$-closed primary
primary states of theory (1) into $\Gz^{(2)}$-closed primary states of
theory (2), and the other way around, whereas $\hat\cA_{-1}$ maps
$\Gz^{(1)}$-closed primary primary states of theory (1) into
$\Qz^{(2)}$-closed primary states of theory (2), and the other way around.
As a consequence, \Qn-closed singular vectors built on \Qn-closed
primaries and \Gn-closed singular vectors built on \Gn-closed primaries
are mapped into each other, between the two topological theories, using
either $\hat\cA_1$ or $\hat\cA_{-1}$.

For no value of $\th\neq0$ the chiral primaries are transformed back
into chiral primaries. Therefore $\hat\cA_\th$ ``destroys''
chiral Verma modules, like the even spectral flow $\cU_\th$ \req{tspfl}.
Under $\hat\cA_1$ and $\hat\cA_{-1}$ the chiral primaries are mapped to
either \Gn-closed or \Qn-closed primaries
annihilated by either $\cG_{-1}$ or $\cQ_{-1}$, respectively,
depending on the specific transformation. Therefore chiral singular
vectors built on \Gn-closed or \Qn-closed primaries are transformed into
singular vectors annihilated either by $\cG_{-1}$ or by $\cQ_{-1}$.

\subsection{Composition Rules}\lvm

The topological spectral flows $\cU_{\th}$ and $\hat\cU_{\th}$, on the one
hand, and $\cA_{\th}$ and $\hat\cA_{\th}$ on the other hand, satisfy the same 
composition rules as their untwisted partners $\cU_{\th}$ and $\cA_{\th}$,
taking into account a ``hat number'' conservation (mod 2), in spite of the
striking differences between the untwisted and the topological spectral flows. 

\vskip .15in
\noi
For the even and the even-odd spectral flows, separately, the rules are

\BE  \cU_{\th_2} \ \cU_{\th_1} =
\hat\cU_{\th_2} \ \hat\cU_{\th_1} = \cU_{(\th_2 + \th_1)} , \qquad
 \hat\cU_{\th_2} \ \cU_{\th_1} =
\cU_{\th_2} \ \hat\cU_{\th_1} = \hat\cU_{(\th_2 + \th_1)} , \EE

\noi
from which one obtains ${\ }\cU_0 = \o1$, ${\ }\cU_{\th}^{-1} = \cU_{(-\th)}$,
${\ }\hat\cU_{\th}^{-1} = \hat\cU_{(-\th)}$, and the relations

\BE \cU_{\th} = \hat\cU_\th \ \hat\cU_0 = \hat\cU_0 \ \hat\cU_\th \ , \qquad
 \hat\cU_{\th} = \cU_\th \ \hat\cU_0 = \hat\cU_0 \ \cU_\th \ ,  \EE

\BE \hat\cU_0=\hat\cU_\th \ \cU_{(-\th)}=\cU_\th \ \hat\cU_{(-\th)} \ . \EE

\vskip .12in
\noi
For the odd and the odd-even spectral flows the rules are

\BE  {\ }{\ }{\ }{\ }{\ }{\ }{\ }\cA_{\th_2} \ \cA_{\th_1} =
\hat\cA_{\th_2} \ \hat\cA_{\th_1} = \cU_{(\th_2 - \th_1)} , \qquad
 \hat\cA_{\th_2} \ \cA_{\th_1} =
\cA_{\th_2} \ \hat\cA_{\th_1} = \hat\cU_{(\th_2 - \th_1)} , \EE

\BE  \cA_{\th_2} \ \cU_{\th_1} =
\hat\cA_{\th_2} \ \hat\cU_{\th_1} = \cA_{(\th_2 - \th_1)} , \qquad
 \hat\cA_{\th_2} \ \cU_{\th_1} =
\cA_{\th_2} \ \hat\cU_{\th_1} = \hat\cA_{(\th_2 - \th_1)} , \EE

\BE  \cU_{\th_2} \ \cA_{\th_1} =
\hat\cU_{\th_2} \ \hat\cA_{\th_1} = \cA_{(\th_2 + \th_1)} , \qquad
 \hat\cU_{\th_2} \ \cA_{\th_1} =
\cU_{\th_2} \ \hat\cA_{\th_1} = \hat\cA_{(\th_2 + \th_1)} , \EE

\noi
from which one obtains ${\ }\cA_{\th}^{-1} = \cA_{\th}$,
${\ }\hat\cA_{\th}^{-1} = \hat\cA_{\th}$, as well as the relations

\BE \cU_{\th} = \cA_{\th} \ \cA_0 = \hat\cA_{\th} \ \hat\cA_0 =
\cA_0 \ \cA_{(-\th)} = \hat\cA_0 \ \hat\cA_{(-\th)}\ , \EE

\BE \hat\cU_{\th} = \hat\cA_{\th} \ \cA_0 = \cA_{\th} \ \hat\cA_0 =
\hat\cA_0 \ \cA_{(-\th)} = \cA_0 \ \hat\cA_{(-\th)}\ , \EE

\BE \hat\cU_0=\hat\cA_\th \ \cA_\th = \cA_\th \ \hat\cA_\th \ , \EE

\BE \cA_{\th} = \cU_{\th} \ \cA_0 = \hat\cU_{\th} \ \hat\cA_0 =
\cA_0 \ \cU_{(-\th)} = \hat\cA_0 \ \hat\cU_{(-\th)} =
\hat\cA_{\th} \ \hat\cU_0  = \hat\cU_0 \ \hat\cA_{\th} \ , \EE

\BE \hat\cA_{\th} = \hat\cU_{\th} \ \cA_0 = \cU_{\th} \ \hat\cA_0 =
\hat\cA_0 \ \cU_{(-\th)} = \cA_0 \ \hat\cU_{(-\th)} =
\cA_{\th} \ \hat\cU_0  = \hat\cU_0 \ \cA_{\th} \ , \EE

\BE \cA_0 = \cA_{\th} \ \cU_{\th} = \hat\cA_{\th} \ \hat\cU_{\th} =
\cU_{\th} \ \cA_{(-\th)} = \hat\cU_{\th} \ \hat\cA_{(-\th)}\ , \EE

\BE \hat\cA_0 = \hat\cA_{\th} \ \cU_{\th} = \cA_{\th} \ \hat\cU_{\th} =
\hat\cU_{\th} \ \cA_{(-\th)} = \cU_{\th} \ \hat\cA_{(-\th)}\ .\EE

We see that the odd and the odd-even spectral flows $\cA_{\th}$ and
$\hat\cA_{\th}$ generate the even and the even-odd spectral flows
$\cU_{\th}$ and $\hat\cU_{\th}$, and therefore they are the only
fundamental topological spectral flows. Observe that $\cU_{\th}$ and
$\hat\cU_{\th}$ commute with each other, as well as with themselves,
whereas $\cA_{\th}$ and $\hat\cA_{\th}$ do not commute with each other,
neither with $\cU_{\th}$ and $\hat\cU_{\th}$, nor with themselves.

\section{Conclusions and Final Remarks}\lvm

In this paper we have analyzed in much detail the spectral flows on the
N=2 Superconformal algebras. For the Antiperiodic NS algebra and the
Periodic R algebra there are two spectral flows: the ``usual'' $\cU_\th$,
written down by Schwimmer and Seiberg \cite{SS}, and the spectral flow
$\cA_\th$, written down by the author and Rosado \cite{BJI4}, which is
an involution quasi-mirror symmetric to $\cU_\th$. We have shown that
the spectral flow $\cA_\th$ is odd and generates the spectral flow
$\cU_\th$, which is even. Therefore only $\cA_\th$ is a fundamental spectral
flow.

For the twisted Topological algebra we have found four different spectral
flows. Two of them act inside the same topological theory: we denote them
also as $\cU_\th$ and $\cA_\th$ because they satisfy the same composition
rules as their untwisted partners. The other two spectral flows
interpolate between the two topological theories corresponding to the two
twistings of the Antiperiodic NS algebra: we denote them as
$\hat\cU_\th$ and $\hat\cA_\th$. They satisfy the same composition rules as
$\cU_\th$ and $\cA_\th$, up to a ``hat number'' conservation (mod 2), and
both have mixed even and odd properties.

The even and the even-odd topological spectral flows $\cU_\th$ and
$\hat\cU_\th$ have been considered in the literature before, but not
analyzed properly. The odd and the odd-even topological spectral flows
$\cA_\th$ and $\hat\cA_\th$, which are involutions, have never appeared
in the literature before. They generate the even and the even-odd
topological spectral flows, and therefore they are the only fundamental
topological spectral flows.

We have analyzed the main properties of all the spectral flows.
We have discussed the properties for general values of $\th$, as well as
the properties for the most interesting values of $\th$, for which
primary states are mapped to primary states (and singular vectors to
singular vectors). Whereas the even and the odd spectral flows $\cU_\th$ and
$\cA_\th$ have quasi-mirrored properties acting on the states of the NS
algebra and the states of the R algebra, the even and the odd topological
spectral flows $\cU_\th$ and $\cA_\th$ have drastically different properties
acting on the states of the Topological algebra. The even-odd and the
odd-even spectral flows $\hat\cU_\th$ and $\hat\cA_\th$ are very similar
to the topological odd and even spectral flows $\cA_\th$ and $\cU_\th$,
respectively.

Finally, we would like to emphasize that the topological spectral flow
transformations $\hat\cU_{\pm1}$ and $\cA_1$, denoted simply as $\cA$, are
``universal'' in the sense that they transform all kinds of
topological primary states and topological
singular vectors back to primary states and singular vectors, mapping
chiral primaries to chiral primaries. All other spectral flow
transformations, except the trivial ones and the mirror map, either do not
map primary states to primary states (the most general situation),
or they do it under restricted conditions.
For example, the topological spectral flows $\cU_\th$ and $\hat\cA_\th$
do not map topological chiral primaries back to chiral primaries for any
values of $\th \neq 0$ (they ``destroy'' topological chiral Verma modules).
Observe that there are no universal transformations for the untwisted
spectral flows, but only for the topological spectral flows $\hat\cU_\th$
and $\cA_\th$.
The case of the NS algebra is the most restricted one: only primary states
that are chiral or antichiral can be mapped back to primary states
(using different spectral flow transformations in each case). As a
result no NS singular vectors can be mapped back to NS singular vectors,
other than the mirrored ones, using the spectral flows, since there are
no chiral NS singular vectors built on chiral primaries, neither
antichiral NS singular vectors built on antichiral primaries.

\centerline{\bf Acknowledgements}

I would like to thank J.I. Rosado for many fruitful discussions and
A.N. Schellekens
for valuable comments and for carefully reading the manuscript.

\end{document}